IAC–23–E2,4,7,x80409

# The Evolution from Design to Verification of the Antenna System and Mechanisms in the AcubeSAT mission

**Panagiotis Bountzioukas[1]**
mpountzi@math.auth.gr

**Georgios Kikas[1]**
kikasgeorg@ece.auth.gr

**Christoforos Tsiolakis[1]**
tsiolakis@ece.auth.gr

**Dimitrios Stoupis[1, 2]**
dstoupis@auth.gr

**Eleftheria Chatziargyriou[1]**
elefthca@math.auth.gr

**Prof. Alkis Hatzopoulos[1]**
alkis@ece.auth.gr

**Vasiliki Kourampa-Gottfroh[1]**
vkourampa@meng.auth.gr

**Ilektra Karakosta-Amarantidou[3]**
ilektra.karakostaamarantidou@unipd.it

**Aggelos Mavropoulos[1]**
mgangelos@meng.auth.gr

**Ioannis-Nikolaos Komis[1]**
komisioann@meng.auth.gr

**Afroditi Kita[1]**
afrodikita@ece.auth.gr

**David Palma[4]**
David.Palma@esa.int

**Loris Franchi[4]**
Loris.Franchi@ext.esa.int

[1]: Aristotle University of Thessaloniki, Thessaloniki, Greece
[2]: Center for Mind/Brain Sciences (CIMeC), University of Trento, Rovereto, Italy
[3]: Dipartimento di Ingegneria dell'Informazione, Università degli Studi di Padova, Padova, Italy
[4]: European Space Agency

## Abstract

AcubeSAT is an open-source CubeSat mission aiming to explore the effects of microgravity and radiation on eukaryotic cells using a compact microfluidic Lab-on-a-Chip (LoC) platform. It is developed by SpaceDot, a volunteer, interdisciplinary student team at the Aristotle University of Thessaloniki and supported by the "Fly Your Satellite! 3" program of the European Space Agency (ESA) Education Office. The scientific data of the mission is comprised of microscope images captured through the on-board integrated camera setup. As the total size of the payload data is expected to be close to 2GB over 12 months, a fast and efficient downlink fulfilling the restrictive power, cost and complexity budgets is required.

Currently, there is no open-source communications system design which fully supports these specific constraints, so we opted to develop our own solutions. The antenna system underwent multiple iterations as the design matured, a process highly aided by the feedback received from the European Space Agency (ESA) experts. The final communications system configuration consists of an S-band microstrip antenna operating at 2.4GHz and a Ultra High Frequency (UHF) deployable antenna, for the payload data and Telemetry & Telecommand (TM&TC) respectively, both in-house designed.

In this paper, we will present AcubeSAT's antenna system iterations that span over 3 years, as well as the rationale and analysis results behind each. The development decisions will be highlighted throughout the paper in an effort to aid in the future development of such a low-cost CubeSat mission communications system.

## 1. Introduction

### 1.1 Mission description

SpaceDot designed the AcubeSAT mission [1], an open-source space biology mission [2] with a twofold goal: First, to probe gene expression changes in eukaryotic cells under the microgravity and radiation effects present in a Low-Earth Orbit. Second, to introduce a LoC platform as a modular means of conducting low-cost, high-throughput Space Biology research. More specifically, *Saccharomyces cerevisiae* cells, one of the best-studied model organisms [3, 4], will be analyzed for a cumulative in-orbit duration of over 7 months, aiming to study fluctuations in the expression of at least 100 genes [5, 6]. To satisfy the needs of the AcubeSAT mission, a 2U experiment platform employing multiple subsystems has been designed [7], as shown in Figure 1.

The microfluidic chip functions as a multiplexing culture device. It consists of 3 sections, each used in-orbit for a distinct experiment run at 3 different timepoints. Live-cell imaging is carried out using a custom microscopy system. Imaging is performed with a temporal resolution of 20 min to perform large-scale analysis of the yeast proteome and a spatial resolution of 100 $\mu$m to resolve individual chambers of the microfluidic chip. The payload data consists of 8-bit 3 color channel images with a resolution of 1280 × 1024 pixels compressed with a compression rate of 2. Since a temporal resolution of at least 20 min is required and as the experiment's duration is 72 hours, a total of $\frac{72h}{20min}$ = 216 images will be downlinked in each run of the experiment which amounts to about 0.425 GB of data. The total number of conducted sub-experiments will be 3, hence the total amount of data to be downlinked during the mission is 1.274 GB [8]. Due to the encoding scheme employed [9], the data field amounts to 3952 out of 5184 bits which is the size of the transmitted codeword. Thus, in order to downlink 1.274 GB of useful data, 1.671 GB need to be transmitted in total. An average communication window of 458.86 $\frac{s}{day}$ is calculated in [8] for a worst-case scenario of LTAN 11. Fulfilling these mission requirements [10] required a communications system that could therefore support a data rate of ≥ 184.8 kbps over a 12-month mission lifetime.

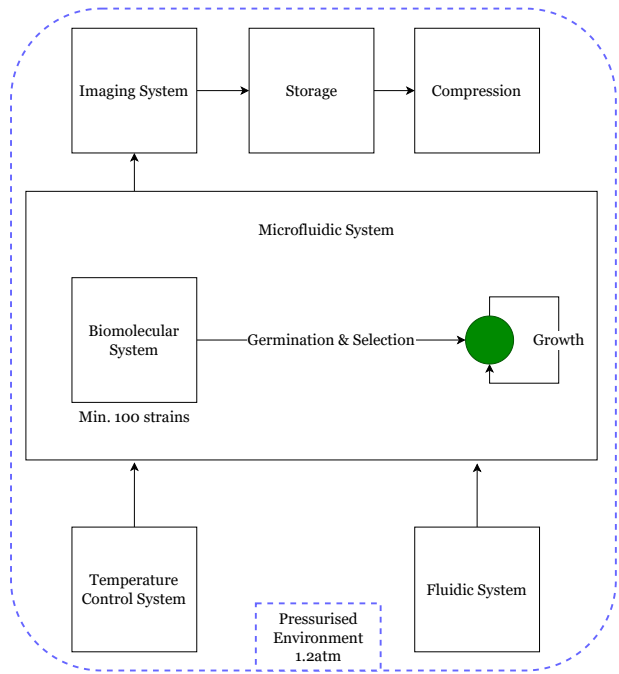

Fig. 1: High-level functions of the AcubeSAT Payload [7]. *S. cerevisiae* cells are housed inside the chambers of a PDMS microfluidic chip. Temperature is regulated for in-orbit cell growth and a set of fluidic components is used for on-demand culturing. The imaging system serves as the data acquisition module for the on-board experiment.

### 1.2 Preliminary antenna system design

The initial proposal for the accommodation of the link and data rate requirements was to utilize two different bands for uplink and downlink, in the radioamateur spectrum. The choice to use this part of the spectrum was attributed to the team's goal to interact with the community and its accessibility, especially for a student team. The downlink would occupy a frequency band, that would provide a high exploitable bandwidth (2.4 GHz) for higher data rate performance, as it concerns both the experiment images and telemetry. On the other hand, the uplink would operate at 1.26 GHz. This initial design decision was made by the team in order to avoid lower bands, such as the UHF, due to overcrowding and the fact that it would require deployable antennas which pose reliability risks.

Furthermore, in order to optimize space usage on the CubeSat outer structure and achieve more solar cell coverage, the team opted for a dual-band patch antenna design. Essentially, it was a microstrip antenna with two square patches, one inscribed inside the other and separated with slots, each resonating at a different frequency. Also, to achieve circular polarization, the corners of both the outer and inner patches were truncated. This design was based on relevant literature such as [11] and can be better visualized in Figure 2.

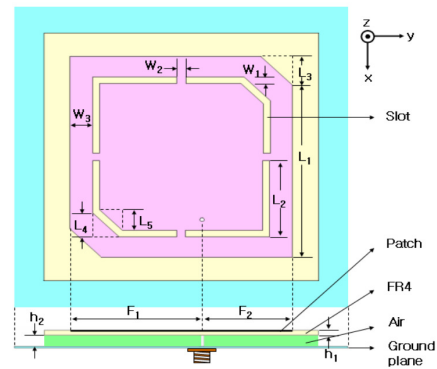

Fig. 2: Dual-band patch antenna design [11, Figure 1].

Although this design adequately served the mission goals, it was discarded before entering Phase C of the mission due to some major disadvantages. It is evident that the use of a directional antenna for all mission operations would pose a strict requirement on the ADCS, and communication with the spacecraft would be unstable in the case of ADCS failure. Moreover, the use of such an antenna would induce unnecessary complexity in the COMMS subsystem design, requiring additional compon-ents such as circulators and dividers to accomodate both frequency bands.

As a result, the team re-evaluated the use of deployable antennas, and the TT&C subsystem architecture evolved to its current state: one S-Band patch antenna for the mission imaging data downlink (section 2) and one deployable UHF antenna for TM&TC (section 3).

## 2. S-Band Patch Antenna

The design of our microstrip (patch) antenna, shown in Figure 3, is based on previous work by Chaouki et al. [12] and Jamlos et al. [13]. The selection of the antenna design and operating frequency is in the same vein as the original dual patch alternative. To minimize Faraday's rotation experienced by a linearly polarized signal through the atmosphere, we opted for circular polarization. The resonance frequency range was set at 2.4-2.45 GHz, with a minimum gain of 5 dB and a HPBW of at least 80°. The primary challenge during the design process was achieving a relatively high gain with a wide beamwidth, to optimize the COMMS power budget without imposing strict pointing requirements on the ADCS subsystem.

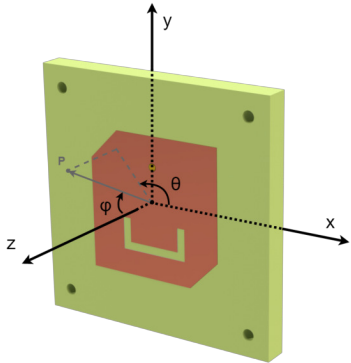

Fig. 3: 3D render of the patch antenna's final design, showing the frame of reference used in the calculations.

A mirror version of the antenna's design will be used in the parabolic reflector of our ground station for the payload data communication link, thus matching the polarization of the one on-board AcubeSAT.

### 2.1 Simulation results

To achieve the desired characteristics, we optimized the designs from [12, 13] until we could meet our requirements in simulations. We aimed to create an inherently circularly polarized patch antenna by truncating the corners and perturbing the design in specific locations with respect to the feed. This approach generates the necessary equal amplitude and in-phase quadrature modes on the antenna.

The configurations responsible for creating a circular polarization operate on the principle of detuning degenerate modes of a symmetrical patch by perturbation segments [14].

In addition to the shape characteristics, selecting the appropriate substrate material posed a further challenge. After careful consideration, we chose FR-4 due to its wide availability and low cost. The substrate was designed to be multi-layer with a total thickness of 4.5mm to achieve a larger bandwidth with higher gain.

The simulation of the final design gives a gain of 4.5 dB (slightly below 5 dB, but within an acceptable margin), and a HPBW of 95°, at 2.43 GHz. The radiation pattern is illustrated in Figure 4. Additionally, the axial ratio remains below 3 dB throughout the range of 2.4-2.45 GHz, ensuring adequate circular polarization.

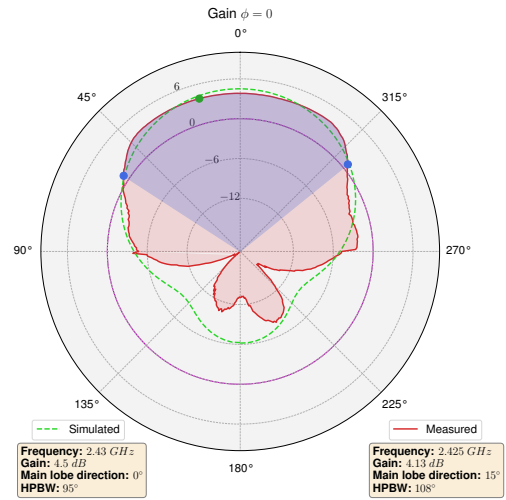

Fig. 4: Simulated and measured radiation patterns of the patch antenna. The ring values are measured in dB. The two dots defining the shaded area depict the HPBW start and end points, while the dot in-between indicates the main lobe direction.

### 2.2 Anechoic chamber measurements

Following the finalization of the design, we conducted a characterization of an in-house built antenna in the anechoic chamber of the Department of Electrical and Computer Engineering at AUTh. The constructed antenna achieved an HPBW of 104°, a gain of 4.13 dB, both at 2.43 GHz, with the radiation pattern shown in Figure 4. Moreover, the axial ratio was measured to be less than 3 dB in the range of 2.42-2.49 GHz, illustrated in Figure 5, closely matching the design specifications of operation within the 2.4-2.45 GHz range.

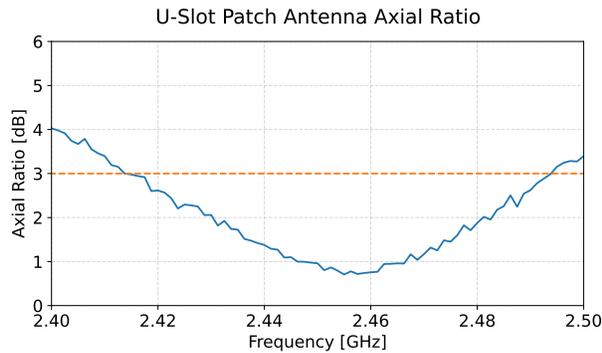

Fig. 5: Measured axial ratio of the antenna.

### 2.3 In-house construction

To achieve the multi-layer design with a 4.5mm thickness, we laminated three distinct FR-4 layers using a two-part liquid epoxy adhesive. The first layer had a pre-sensitized (photosensitive) copper pour on one side and on the other bare FR-4. The middle layer consisted of a bare FR-4 and the back layer had a copper pour on one side, with the bare side laminated with the middle layer. The design of the antenna was imprinted on the pre-sensitized copper board and after developing the imprint, the rest of the unwanted copper was etched away using ammonium persulfate as the etching agent. Parts of the procedure are shown in Figure 6.

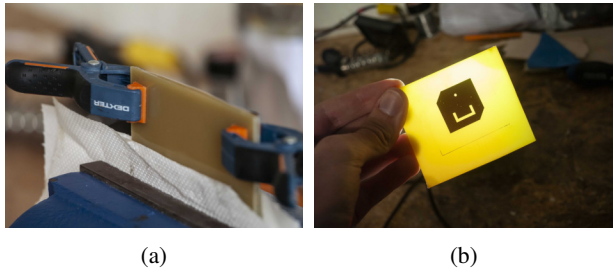

(a)          (b)

Fig. 6: In-house construction of the patch antenna. a) Lamination of the bottom and middle layer, b) Imprinted top layer. Figures from [15].

### 2.4 Detailed reports and supporting material

For detailed information on the design iterations, construction procedure, and comprehensive analysis of the measured results, refer to the relevant report [15] and repository [16].

## 3. UHF Antenna — Deployment Mechanism

The strict pointing requirements incurred by the use of an S-Band directional antenna, despite being considered necessary for the mission imaging data downlink, led the team to opt for the use of an antenna characterized by a radiation pattern close to omni-directional, to ensure the independence of critical TM&TC from high-risk flawless ADCS function. Furthermore, due to both TM&TC's low data volume, a lower datarate would be required, thus a frequency band on the lower end of the spectrum would suffice. The chosen frequency band was the 70cm UHF band, which is also the most commonly used in LEO CubeSat missions [17]. The length of the antenna elements required to operate in this band in conjunction to the limitations imposed on the envelope of the CubeSat structure by the deployers, as in [18] for example, make the use of a deployment mechanism inevitable.

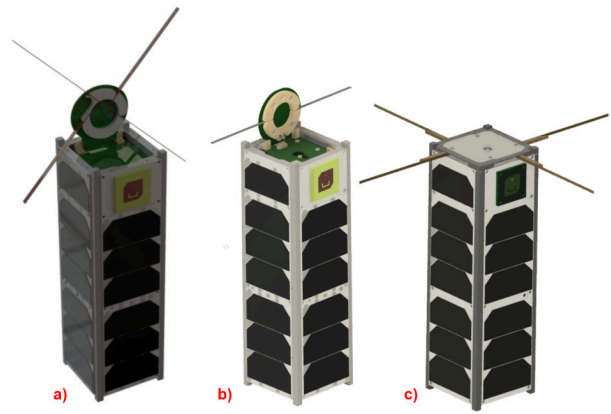

Fig. 7: AcubeSAT bus with ADM integrated a) v0, b) v1 and c) v2 (current).

In general, the UHF antenna and deployment mechanism had to abide by certain RF criteria that the mission demanded. More specifically, the antenna had to have minimal pointing losses regardless of the spacecraft orientation and impose minimal polarization losses to the link. A turnstile antenna, often referred to as a crossed dipole antenna, was chosen in order to fulfill these link related requirements. On the other hand, the deployment mechanism underwent multiple iterations, before the team concluded to the current model. This section will illustrate the evolution of the UHF antenna and highlight the lessons learned throughout the process.

### 3.1 Initial design based on UPSat's mechanism

The first approach the team took was using the UPSat ADM as a basis, tailoring it to the mission's needs, due to the fact that it was open-source and flight-proven [19]. Another asset of this design is its modularity, given that it wouldn't require any modifications on the spacecraft frame structure. This mechanism, designed by Libre Space Foundation for the UPSat mission, was used to de-ploy two monopoles, a VHF and a UHF one.

The main premise of this design is using a deployable lid mounted on the +Z face of the CubeSat, around which the antenna elements are held folded before commissioning, and deploy exploiting their accumulated spring force. The lid is connected to the burner circuit PCB, which is mounted on the frame, via two hinges and a rod running through them. A stopper is also utilized, in order to constrain the lid's movement to up to 90° with respect to the +Z face. In addition, two torsion springs are wrapped around the rod, to provide the tension for the lid to be deployed. Furthermore, the lid is secured into its stowed configuration with two melt lines tied via a hitch knot, connecting the burner circuit PCB and a support on the lid. The two melt line redundancy ensures that the mechanism will endure the mechanical loads during launch and stay stowed, even in the case one of the melt lines breaks.

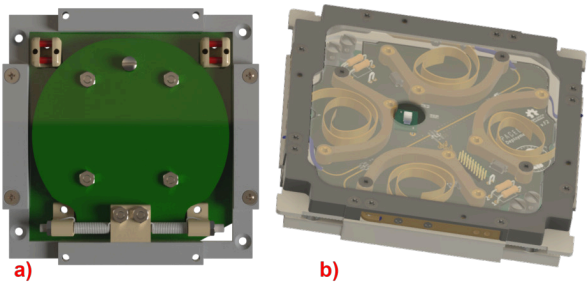

Fig. 8: Deployment mechanism a) v1 & b) v2.

One of the first challenges with this approach was to modify the design from housing two monopoles into one that can house two crossed dipoles for a turnstile antenna. More specifically, the antenna holder, around which the antenna elements are curled, would be modified to accommodate the two orthogonal dipoles in a manner that no plastic deformation could be imposed on the elements. It is also noteworthy, that the system's geometry posed limitations on housing two typical dipoles on the desired orthogonal configuration. Consequently, as seen in Figure 9 (right) in one of the dipoles the elements aren't collinear with a small deviation, but are parallel to stay in-line with the crossed dipole specification.

In order to account for this deviation from an ideal turnstile antenna, simulations for the radiation pattern and axial ratio were ran to verify that this configuration satisfies the RF requirements. The simulation results, presented in detail in the relevant team report [20], ensure that this antenna design performs adequately. Another key takeaway is that a feeding circuit was designed and integrated as well on the lid PCB, as a matching network between the unbalanced RF signal used by the COMMS

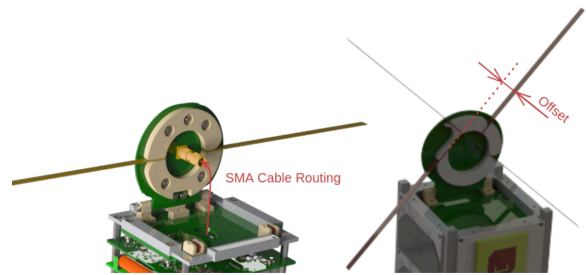

Fig. 9: *Left*: ADM v1 cable routing, *Right*: ADM v0 dipole element offset.

Board and the antenna element pairs that required currents of equal magnitude and a $\frac{\pi}{2}$ phase offset between them.

However, several problems existed with this design:

- The aforementioned matching network on the deployable lid required a four cable interface, attached directly on the inner side of the lid. Also the space limitations on this side of the lid, given the antenna holder was also present, posed a challenge in integrating the matching network efficiently.

- The fact that 4 antenna elements, perpendicular to each other had to be deployed in this manner, posed serious reliability issues. The risk of one of the lid's bottom elements being hindered by the CubeSat structure during deployment was major, given the fact that the ADM is a single point of failure in the AcubeSAT mission. This is more evident in Figure 9 (right) where the lack of major clearance between the bottom elements and the structure is illustrated.

- An additional important deterring factor was the risk of damaging the solar cells with the antenna elements. This is attributed to potential oscillations of the elements due to their spring force, during deployment, and the fact that they are co-planar to the solar panels

Given all of the above, the team then compromised on using a single dipole antenna instead of the turnstile configuration in order to be able to utilize the UPSat inspired deployment mechanism design.

This iteration would be compliant with the beamwidth and gain requirements while also increasing the reliability of the deployment. However, using a single dipole would mean that the polarization requirement wouldn't be met, as linearly polarized waves would be produced instead of circularly. Given that the ground segment would be equipped with a circularly polarized helical antenna, only a 3 dB constant polarization loss would be imposed, which according to the link budget analysis [8] was acceptable. A more in-depth analysis of this iteration along with the antenna simulation results can be found in the relevant report [21].

However, regardless of the assumptions made during the system design, issues with this iteration surfaced as well, also following the prototyping done by the team. A mock-up was initially 3D printed to allow for preliminary testing before finalizing the design. The testing that followed indicated issues with the element wounded on the bottom side of the antenna holder to unfold, which was a major drawback. Some considerations were made to modify the antenna holder in such a way that both elements would wind around the top side of the lid, but eventually this design was rejected.

Another driving force for this decision, other than the inherent deployment complications, was the interface of the antenna with the COMMS Board. Although the SMA cable would provide a lossless connection between the two and directly feed the antenna, its size, along with the fact that it had to be mounted on the lid, was sub-optimal with respect to space management, as seen in Figure 9 (left). Additionally, the SMA cable, being mounted on the lid could itself pose risks to the deployment reliability as well, as another moving part in this design could insert unwanted friction forces.

All in all, while the UPSat deployment mechanism proved to perform well in its pilot mission, modifying it to accommodate the turnstile design posed a risk of incorrect deployment for our mission.

## 3.2 In-house ADM development

The next iteration (Figure 10) that was chosen was based on a roll-up configuration that is prevalent among CubeSat antenna deployment mechanisms [22, 23, 24, 25]. A turnstile configuration has been chosen for the antenna in order to satisfy the polarization and omnidirectionality requirements. All the design files for this version of the mechanism can be found on GitLab [26]

The main advantage of the design is its relative simplicity accompanied by deployment consistency.

During the design and development of the mock-up, it was found that successful deployments could be achieved more reliably compared to the previous lid design. The main problems identified during preliminary testing lied on the repeatability of the refurbishment procedure. Routing, knotting and capstan tensioning procedures proved to be particularly hard to replicate, especially between different operators.

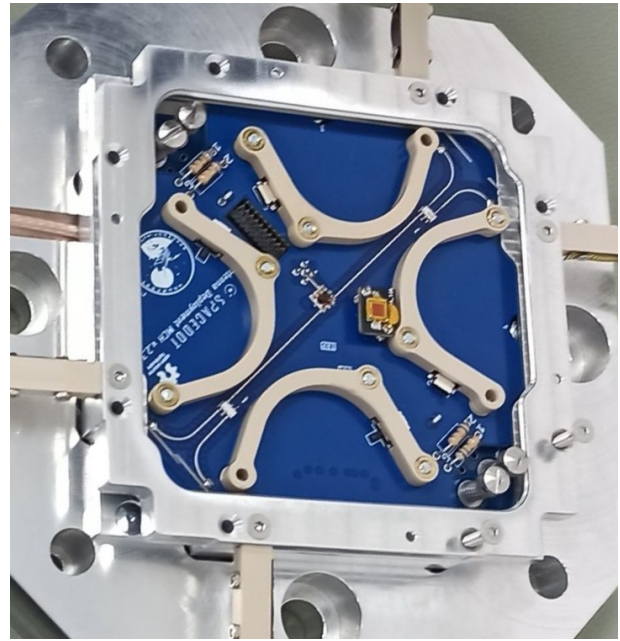

Fig. 10: ADM EQM Design.

### 3.2.1 Physical Architecture

The AcubeSAT ADM physical architecture is shown in Figure 11. Its main components are listed below:

*Top/Filler PCB* - This is the main component of the mechanism which accommodates the RF circuitry, the deployment switches, the CSS and the antenna holders.

*Bottom PCB* - This board, mounted on the Team Frame, serves as the electrical interface with the rest of the satellite and houses the burner circuit.

*Team Frame* - The mechanical interface of the ADM with the PC/104 PCB stack which is directly below the mechanism.

*Cover Frame* - A structure placed on top of the Satellite Frame to serve as a fixture for the doors' axes while adding multiple secure points for the whole assembly.

*Satellite Frame* - Aluminium structure located in between the Team Frame and the Cover Frame. Part of the AcubeSAT nanosatellite body structure.

*Thermal Knifes* - Two sets of two resistors, each controlled by the burner circuit, through which the melt lines are routed and cut during deployment.

*Antenna Elements* - The elements of the turnstile antenna. Stainless steel was used for the EQM, since the proposed beryllium copper alloy was difficult to procure in a timely manner. It was decided that C17200 berillium alloy will be procured for the FM. This material is preferred for being both light and stiff in respect to other metals. Any bending or plastic deformation caused while in stowed state, is considered to be more difficult to manifest in respect to other alloys [27]. Such alloys also have legacy in similar mechanisms.

- *Doors* - PEEK structures on which the antenna elements are mounted. Torsion springs are placed in the doors' rotation axis to facilitate the actuation while also applying tension to the dyneema. PEEK was purposefully selected to be different from the frame to any avoid cold welding and low thermal expansion/friction effects.

- *Torsion Springs* - Provide the tension for the doors and as a consequence the mounted antenna elements, to be deployed.

- *Pivots* - Metallic reverse U-shaped pins placed on Top PCB for dyneema routing.

- *Antenna Holders* - Structures which keep the four antenna elements folded when in stowed position. They incorporate inserts thus introducing threads for parts interfacing.

- *Melt lines* - Dyneema lines which hold the doors closed, melt and get cut when the deployment sequence is initiated. The diameter was chosen based on the maximum load plus margin and through extensive testing (multiple deployments).

- *Capstan Screws* - The four melt lines are tied to 4 capstan screws using a Midshipman's hitch knot (#1855 in *The Ashley Book of Knots* [28]). Compression springs are placed below them (Figure 15). By further screwing the capstans, more tension is applied to the lines. The team followed a specific procedure to ensure that the tension would be the same along all 4 melt lines. The detailed method can be found in [29] or [30].

- *Cover* - The top enclosure of the mechanism, which constrain the antenna elements on the Z axis. There is a hole near the geometric centre for the CSS. During subsystem and later during system level environmental tests, another hole to plug the system level harness is incorporated on to the cover. The FM cover will not share this feature, to avoid any thermal related issues.

- *CSS Board* - A daughter board containing part of the coarse sun sensor (SFH2430-Z photo-diode), required by the ADCS for the mission, is soldered on the Top PCB. The mounting is realised via the direct soldering of headers on the top PCB.

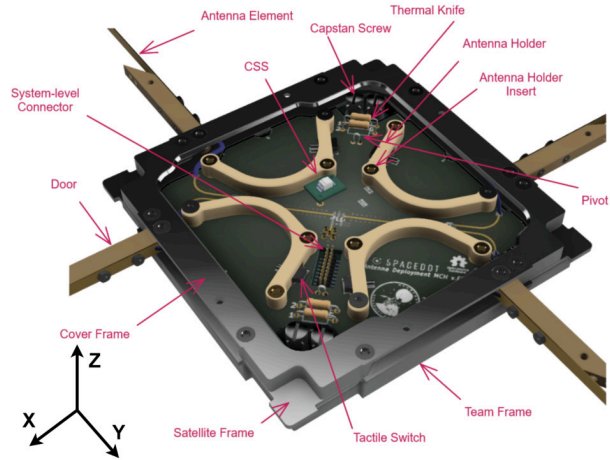

Fig. 11: The physical architecture of the AcubeSAT ADM. The PEEK cover on top of the mechanism is hidden and the Bottom PCB is not visible.

The dimensions of the ADM are 100 x 100 x 5.5mm (XYZ) where Z axis is the height measured from the satellite frame. The total height stands at 15.6mm.

Table 1: AcubeSAT ADM: Mass Budget.

| Component | Mass in gr |
|---|---|
| PEEK Antenna Holders | 5 |
| Aluminium Frames (Team, Satellite and Cover) | 74 |
| PEEK Cover* and Doors | 14 |
| Top and Bottom PCBs | 41 |
| Bolts, Springs, Nuts, Elements | 33 |
| Margin | 10 |
| **Total Mass** | **177*** |

[*]: Incorporating improvements (subsection 3.4) will increase overall mass by 10gr.

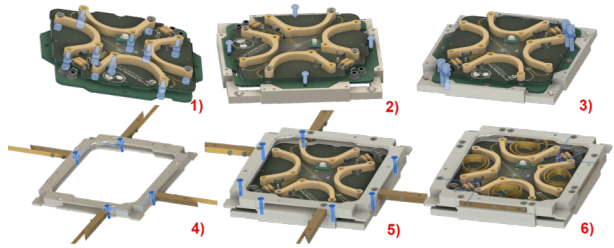

Fig. 12: Assembly procedures for the AcubeSAT ADM. Mounting of the PEEK cover on top of the mechanism is not shown.

The assembly of the AcubeSAT ADM is split into the following sub-assembly routines, better illustrated in Figure 12:

1. A sub-assembly consisting of the antenna holders and Top and bottom PCBs. After placing the board-to-board cable assembly, bolts are used to fix together the parts. The bolts pass through the bottom PCB and the Top PCB until they are screwed on the antenna holders' threaded inserts. Note that, threadless spacers are used to maintain the required vertical distance between the two PCBs. The thermal knife through-hole resistors are later soldered into position. This sub-assembly is then mounted on the Teamframe using one bolt for each side.

2. Four bolts are placed as shown in Figure 12 to secure the previous sub-assembly to the *Team Frame*.

3. Compression springs are placed together with the capstan screws onto the bottom PCB's dedicated mounting holes.

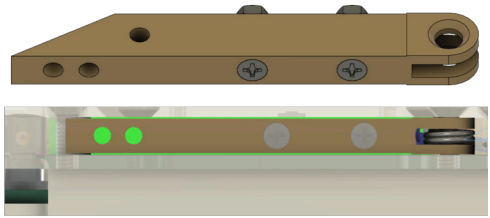

Fig. 13: One of the four doors on which an antenna element is mounted. Clearance was added (denoted with green) to avoid friction between the door and the Cover and Satellite Frames.

4. A cable is soldered on the antenna elements, in between the mounting holes. The other end of the cable is left to be soldered during step no.5 on to the pads where the ministrip ends (elements-PCB interface). Each antenna element is placed on the doors and held in place with the use of two pairs of bolts and nuts. A provision concerning the friction between the door (Figure 13) and the frame was made by incorporating an intrusion to the door surface. Clearance was smaller in the door axis to avoid any possible angular inclinations that would result in the element being lodged in between its neighboring surfaces. The doors are then placed in between the satellite frame and the cover frame by placing the axes' bolts and torsion springs. It should be noted that the door nuts are the only ones used in the current iteration of the design. Although nuts were initially considered as the main elements for bolt fixturing, they were ultimately discarded in favor of the assembly simplicity and were replaced with either threaded holes or inserts.

5. Sub-assembly no.4 is later screwed on top of the teamframe-PCBs sub-assembly. The other end of the cables on the elements, are being soldered on the pads as described in routine no.1.

6. The melt lines are routed as shown in Figure 14, by connecting them through a pair of holes in each door to the capstan screws. The melt line is initially fixed on each door by performing a knot through the two side holes. The doors are then pushed inside the structure after nesting the elements in the Antenna Holders. To hold the doors in place, thus countering the torsion force, a holding pin which passes through a hole in each door is used. The melt line is then passed by the operator through the pivot A, to the top and bottom of the resistors and through pivot B (Figure 15) to the capstan bolts where it is secured using a Midshipman's hitch knot. Further tensioning of the dyneema is possible by further tightening the capstans. Each melt line is routed through two resistors, each belonging to one of the two thermal knives, to introduce redundancy.

7. The final step of the assembly is the placement of the PEEK cover, fixed by four bolts. In case the operator wishes to reroute a dyneema line after a possible deployment test, the cover along with the previously cut melt lines can be removed and step number no.6 can be repeated.

In earlier iterations of the design the melt line's termination point used to be a simple screw with a mounted tension spring as shown in the development model depicted on the left in Figure 14. The reason for the design choice to change to capstan screws is twofold: Increase the space available for routing and most importantly, provide the operator with the option to apply more or less tension to the melt line by screwing or unscrewing the capstan screw.

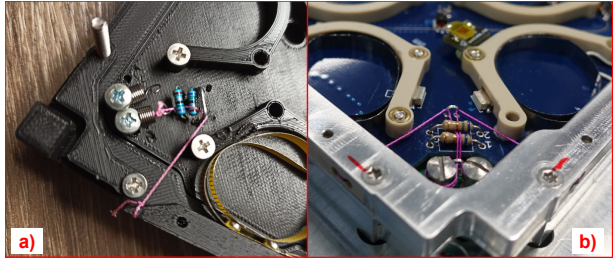

Fig. 14: Melt line routing. a) An earlier development model which utilised regular screws with tension springs mounted. b) The EQM model of the ADM with the updated capstan screw design. The melt lines are knotted using a Midshipman's hitch knot in both edges (the two holes of the door and the capstan screw). The melt line is carefully routed through the thermal knife.

Moreover, in order to safely secure the capstan screw after arming the mechanism, a compression spring applying a normal force is utilized, which is placed between the capstan and the threaded spacer as indicated in Figure 15. The force exerted by the spring to the capstan increases the static friction force which can in turn counter the torque that the melt line applies to the screw, that could result on loosening the termination. Thus, by inserting the vertical tension spring, it is reassured that the melt line won't come loose (Figure 15).

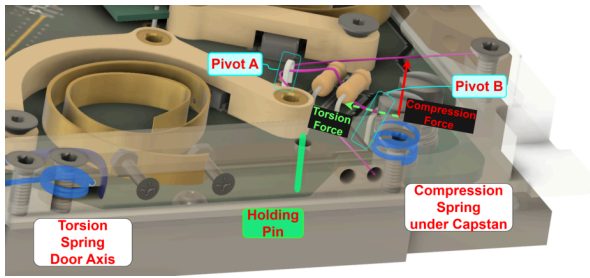

Fig. 15: Capstan melt line termination, mounting of the capstan and spring justification.

### 3.2.2 Electrical Design

The electrical design of the ADM, consists of the RF feeding circuit, the burner circuit and other electrical interfaces specific to the AcubeSAT mission, such as a coarse sun sensor, a board-to-board cable assembly and the umbilical cable connector for the system level AIV activities.

The RF part is housed on the Top PCB and uses an MMCX connector on its bottom side to interface with the on-board RF transceiver. To be able to feed properly a crossed-dipole antenna, the signal is split and phase-shifted by 90° in a manner that the two perpendicular dipoles are fed with quadrature signals. This is achieved with one SBTCJ-1WX+ power divider after the coaxial connector and two QCN-5D+ power splitters (by Mini-circuits) that also introduce the required phase offset.

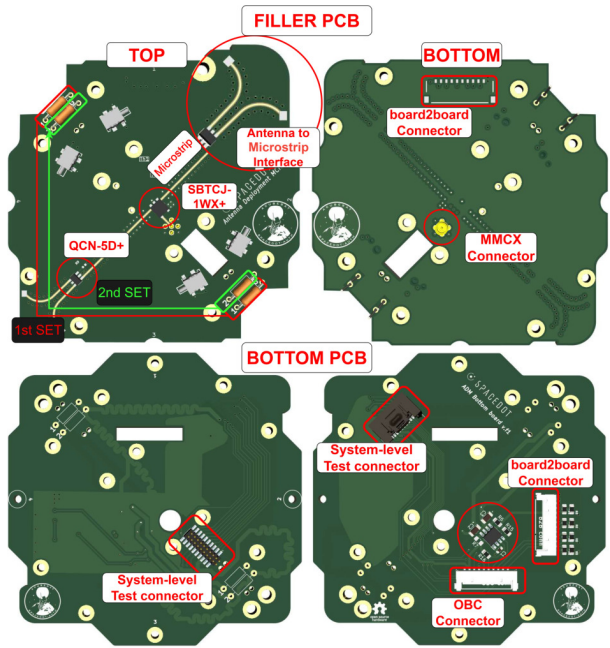

Fig. 16: Electrical Interfaces of the in-house ADM.

Regarding the burner circuit, it is mostly accommodated on the Bottom PCB and it consists of a BD2069FJ-MGE2 power switch IC, that supplies with 5V the dyneema burner resistors (6.8 Ohms, 1/4W) and also provides over-current protection. The selection of the resistors was made after extensive testing both on ambient and lower temperatures (use of a freezer). It was also ensured that the EPS could provide the required power through the regulated 5V channel. The power switch is supplied and controlled externally, and specifically for our mission by the OBC. A cable assembly is used to facilitate the connection. The deployment is detected using four tactile switches held at their off state during the stowed configuration, as they are pressed by the folded antenna elements. Thus, a feedback loop is created between the actuation and detection of the deployment, a procedure executed by an external controller. This control logic is described in detail in the team's CDR design justification document [27].

Redundancy is also another noteworthy characteristic of the AcubeSAT ADM as this reflects on the burner circuit. As mentioned in Section 3.2.1, the mechanism's two thermal knives consist of two resistors, which in fact are supplied by different outputs of the power switch. These outputs can be controlled separately reflecting the double redundancy on the thermal knife supply layer.

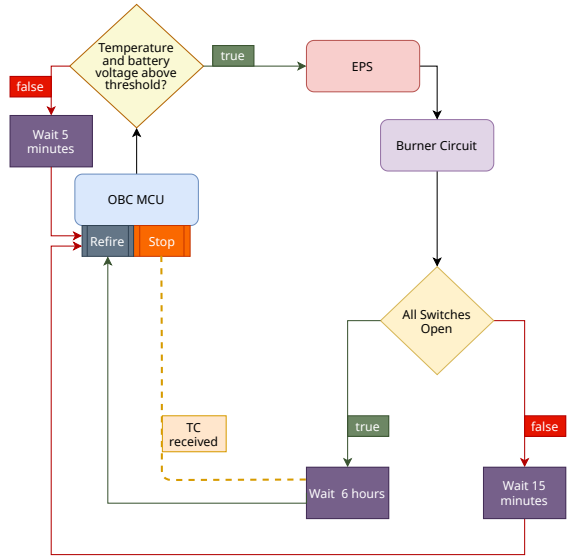

Fig. 17: Deployment feedback logic diagram.

In order to initiate the deployment procedure, which is also illustrated in Figure 17, health checks on the spacecraft's critical parameters during commissioning take place (e.g. battery voltage). Once the conditions prove adequate, the OBC enables the power switches channels consecutively, heating up a burner resistor set (one on each side Figure 16). Once the melt lines eventually burn, the elements are released and the tactile switches' state changes. Concurrent firing of the resistor sets is not preferred, since space debris might formulate in the form of cut line residue (in between the resistor sets).

If all switches appear as open, a deployment is attempted anew after 6 hours, unless a telecommand is received in the meantime, which serves as an assurance to the OBC that there were no false positive deployment signals. In order to account for the scenario where a TC is received while only a subset of the elements are deployed, a separate 24-hour timer is employed which forces the burning circuit to actuate. This timer can later be overwritten if the operators deem it necessary.

Otherwise, it is re-attempted after 15 minutes, if the conditions are favorable. This was chosen in view of avoiding prolonged deployments, that could deteriorate the resistors and consequently decrease their efficiency.

### 3.2.3 Cost Analysis

The overall costs concerning the EQM are considered to be on the low-end of available commercial options. The breakdown of the costs can be found in Table 2.

Table 2: AcubeSAT ADM: Cost per Component.

| Component | Cost in € |
|---|---|
| PEEK Antenna Holders | 120 |
| Aluminium Frames (Team, Satellite and Cover) | 620 |
| PEEK Cover* and Doors | 910 |
| Top and Bottom PCBs | 50 |
| Electronic Components | 50 |
| Capstan Screws, Springs and Bolts | 100 |
| Antenna Elements | [*] |
| **Total Cost** | **1850** |

[*]: Already available stainless steel elements were used on the scope of the qualification. This cost corresponds to the the EQM model. Modifications that are described in subsection 3.4, are to be made.

### 3.3 Environmental Testing

An Environmental Testing Campaign with an EQM model of the AcubeSAT ADM design serving as the DUT was conducted from 05/12/2022 to 09/12/2022 at the ESA CubeSat Support Facility, ESEC-Galaxia, Transinne, Belgium [31]. The campaign consisted of Vibration and TVAC testing. The detailed Test Specifications and Test Report can be found in [29, 30, 32]. In this section, we highlight the main takeaway points the team received to improve their design.

#### 3.3.1 Vibration Testing

During Vibration Testing the DUT underwent, on each axis, Quasi-static Acceleration for 1 second and then Random Vibration testing, preceded and followed by Resonance Searches to verify structural integrity by observing minimal frequency and amplitude shifts in the frequency response. It is notable that the testing profiles specified in GEVS [33] were used.

The accelerometers were mounted as seen in Figure 18. Three main measurement accelerometers were used, placed on the corner and the center of the ADM's top cover and on top of the surrounding frame respectively. Two control and one co-pilot accelerometers were also placed on the mechanical adaptor to facilitate the operation of the shaker.

As elaborated in [32], it was clear that the accelerometers' Frequency Responses were almost exactly the same for every Resonance Search performed before and after the Quasi-static Acceleration and Random Vibration sequences, that is, both frequency and amplitude shifts were negligible. Also, after the tests, the DUT functionality remained unimpaired. As a result, the testing process was deemed generally successful.

Table 3: Accelerometer Placement.

| Reference | Type | Location |
|---|---|---|
| C1 | Tri-axial | Adaptor plate |
| C2 | Tri-axial | Adaptor plate |
| M1(CP) | Mini Tri-axial | Adaptor plate |
| M2 | Mini Tri-axial | Corner of PEEK cover |
| M3 | Mini Tri-axial | Centre of PEEK cover |
| M4 | Mini Tri-axial | Frame |

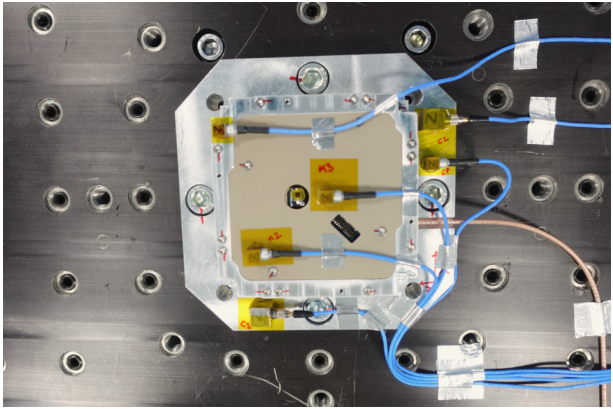

Fig. 18: Accelerometer Placement during Vibration Testing.

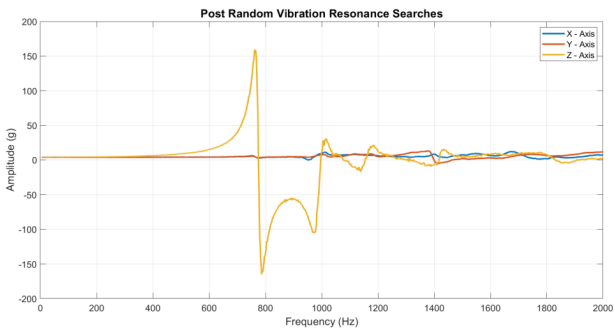

Fig. 19: Data gathered from the accelerometer placed at the center of the PEEK cover during the Post Random Sequence Resonance Searches performed on each axis. An extreme amplification factor of approximately 40 is observed in the Z axis measurement.

However, an observation worth noting from the Z axis test is the extreme amplification factor for the resonance amplitude of the cover. Measurements from the central accelerometer (Figure 19) showcased an amplification factor of approximately 40 during the Z-axis final Resonance Search. This value was a result of:

- Certain protruded inserts on the antenna holders (interface points) that induced height variations between the four mounting points of the cover
- The incorporation of complementary fillets between the cover frame and the cover since there were oscillations between these two parts
- The placement of the actual accelerometer on top of the cover

The observed bending of the cover, that deemed the PEEK cover unreliable, could therefore be a product of any of the errors mentioned, including the vibration tests. The proposed solution is being explained in subsection 3.4.

*3.3.2 TVAC Testing*

Whilst in TVAC testing, the DUT was subjected to a thermal cycling profile which incorporated 1 non-operational and 3 operational cycles during which 3 deployments were attempted:

- An extreme cold case deployment at -25.5°C
- A cold case deployment at -15°C
- A hot case deployment at 50°C

Table 4: Thermocouple Placement.

| Reference | Function | Location |
|---|---|---|
| TC 50 | TRP | Central Reference Point |
| TC 51 | Measurement | PEEK to FR4 |
| TC 52 | Measurement | Aluminium to PEEK #1 |
| TC 53 | Measurement | Aluminium to PEEK #2 |

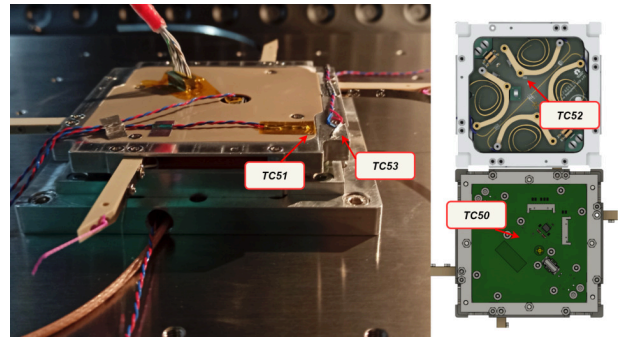

Fig. 20: Thermocouple Placement.

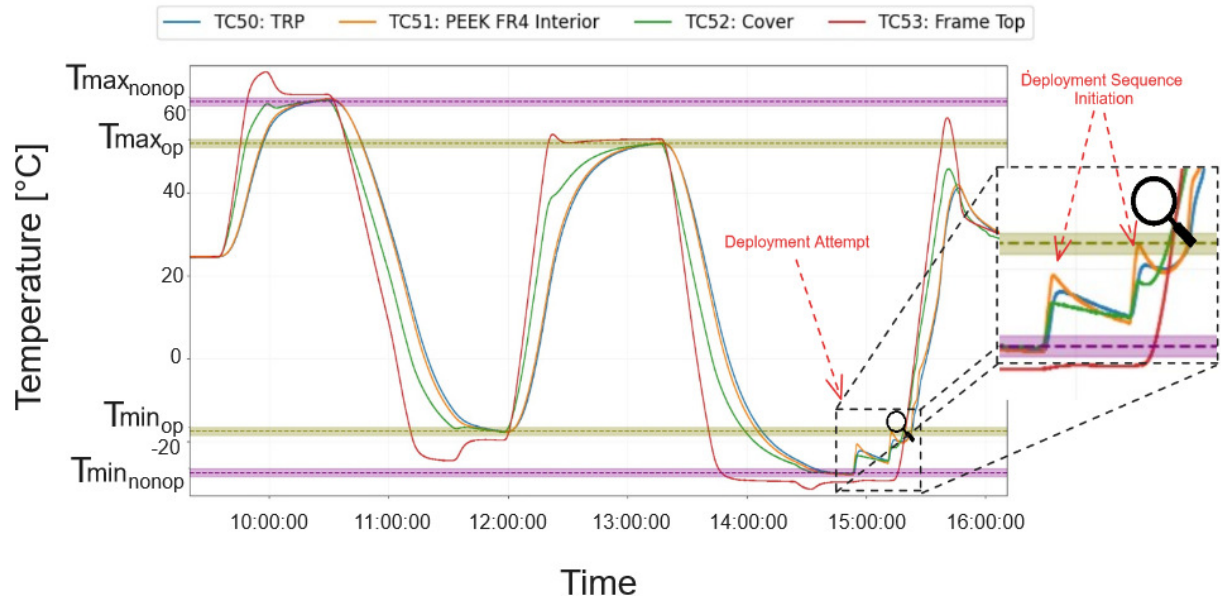

Fig. 21: Extreme Cold Case: Thermocouple Data. The spike observed corresponds to the initiation of a deployment sequence. A second attempt, making use of the redundant thermal knife, was made. All thermocouples measure increased temperatures.

In total, four thermocouples were placed on top of the DUT. TC 50 was placed on the Z- side of the bottom PCB, having a relatively long distance from the highly dissipative resistors. TC 51 was placed on the top PCB, near the center of the board. TC 52 and TC 53 were placed near the PEEK to Aluminum interface (cover and frame respectively) to better observe the conductive interface.

Various minor anomalies were detected during testing (as elaborated in [32]) but no damage was observed due to cyclic expansion and contraction and all circuitry remained operational after testing. Hot and cold case deployments were successful, however, the most important takeaway from the test was that the antenna failed to fully deploy (only two doors opened after two deployment attempts) when tested for the extreme cold case (-25.5°C).

The reason for the failed deployment was that the dyneema lines did not reach their melting point fast enough if not at all. This is also visualised in Figure Figure 22, where we observe the difference in the way dyneema is cut during an ambient temperature test (clean cut) and during the extreme cold case where significant melt residue can clearly be observed. The temperature of the dyneema starts dwelling before reaching said melt point of the lines and it is finally cut after enough power is dissipated for the material to change state. A measured observation of this phenomenon can also be found in [23].

In the extreme cold case the heat dissipated is just below the threshold needed, since after stressing the line with two deployment sequences only two out of four doors were deployed. It can be seen from Figure 21 that all thermocouples - except for the one on the top of the frame - showcase significantly increased temperatures. This led the team to believe that the presumed root cause of this observance is insufficient thermal insulation of the heating elements with the rest of the board.

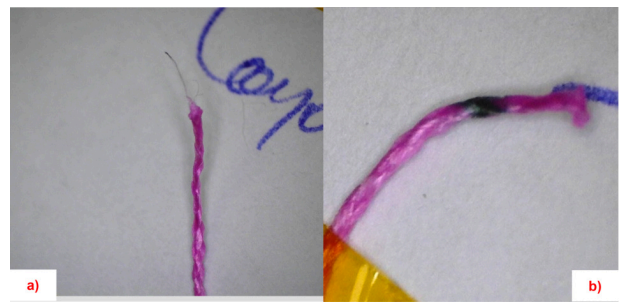

Fig. 22: a) Melt line cut during the coldest case deployment. b) Melt line cut during an ambient temperature deployment.

In particular, it can be observed that for the current iteration of the design, the copper ground plane, having a re-latively small thermal resistance and covering a large surface area, creates an easier heat transfer path to it than the dyneema. That in turn leads to limited heat power being dissipated to the dyneema. An important data point is the fact that even at the extreme cold case, a partial deployment was achieved, while during the cold case of -15°C, the mechanism could be fully deployed. This implies that the power dissipated is close to the threshold needed to cut the dyneema. By further thermally insulating the heating elements from the ground plane, it is expected that more power will be dissipated to the melt line, allowing the mechanism to pass the power threshold required for the deployment.

### 3.4 Possible Improvements

While the environmental test was considered a success, the fact that some non-conformances were reported, imposed the need of defining clear next steps and improvements towards the finalization of the mechanism. The improvements examined in this section not only address the issues observed during the test (better described in Section 3.3.2) but also reveal certain other design paths that were not considered initially due to the tight mission timeline and the fast-paced development that this induced.

- *Cover*: Regarding the high amplification factor of the cover acceleration during Vibration Testing, one potential consideration is switching from PEEK to a more rigid material, such as aluminium (Figure 23 (1)). This will provide enhanced structural integrity and mitigate any potential issues related to resonance amplification in future testing scenarios. However, updated thermal simulations are needed to examine whether the change of material will have any notable effect on the system. The filleted surfaces of the cover frame shall also be removed as an additional amplification mitigation measure.

- *Thermal Insulation*: Due to the known thermal transfer inefficiency problem better described in Section 3.3.2, the action the team opts for is to decrease the overall area of the ground plane, as seen in Figure 23 (4), aiming to focus the produced heat towards the dyneema.

- *Antenna Element - PCB Interface*: In the current design, the antenna elements are directly connected to the Top PCB's microstrip through a soldered single core cable. This interface proved to be prone to failures due to the fact that solder joints reside on a moving component (door). This increased the risk of cable rupture due to fatigue stress. For this reason, the team plans to opt for coaxial cables that connect with the elements using ring terminals and are screwed together on the element doors, which eliminates any soldering issues (Figure 23 (2)). However, the stiffness of the coaxial cables could introduce deployment issues that shall be identified with extensive prototyping.

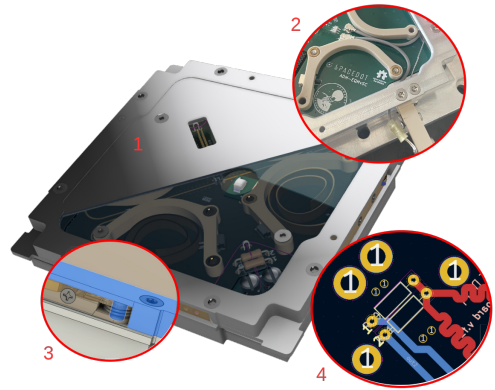

Fig. 23: ADM suggested improvements

- *Resistors to bottom PCB*: Another aspect of the mechanism that could be improved, is the location of the thermal knife resistors. Since their pins pierce both PCBs, once the resistors are soldered, the PCBs are locked together in a sub-assembly. This introduces difficulties to access certain points on the PCB useful during troubleshooting, while it also introduces heat dissipation issues. The proposed design change would be to have the resistors connected to the bottom or the top PCB alone. This would result in a modified melt line routing since the dyneema would also have to traverse a vertical distance, due to the height difference of the PCBs. Since future development is expected, this change would be required to facilitate it.

- *Doors - Frame Mechanical Interface*: Yet another consideration would be to alter the means with which the element doors are mounted on the frame. In the current version, bolts are used as both mounting points and axes of rotation, induced by the torsion spring located around the bolts. In order to avoid possible friction between the door and the frames that could result to deployment failure, these bolts are fastened with a lower torque. Although this decision posed risks regarding the impact of the launch loads on the mechanism, the vibration test proved otherwise. Nevertheless, an iteration where the axes are cylindrical extrusions of the cover frame instead of bolts could eradicate any possible counter-to-the-torsion-spring forces related to increased friction as shown in Figure 23 (3).

## 4. Conclusion

In this paper, we traced the evolution of the AcubeSAT antenna system, a journey driven by the mission's stringent demands for high data rates and reliable TM&TC support. The resulting antenna system consists of an S-Band patch antenna, operating at 2.45 GHz, and a UHF Turnstile antenna working at 436.5 MHz, complemented by a deployment mechanism.

The development of the S-Band patch antenna, drawing inspiration from designs presented in [12] and [13], was outlined. We emphasized the iterative optimization process undertaken to meet the requisite resonance frequency range, gain, and HPBW. Validation of the design was substantiated with measurements obtained from a characterization conducted at the anechoic chamber of AUTh, while the procedure for the in-house construction was presented in summary.

Subsequently, we delved into the extensive journey of the AcubeSAT ADM. Initially, the team explored the adaptation of the UPSat ADM, custom-tailored to meet the specific needs of our satellite. However, the team ultimately decided to pursue an in-house ADM design. We offered valuable insights into the development process, encompassing various design considerations, the physical and electrical architecture of the current ADM iteration, and the wealth of lessons learned during each phase. Furthermore, we presented the outcomes of the environmental testing campaign conducted on the EQM of the current ADM design. Lastly, we explored potential avenues for enhancing the mechanism.

As the project continues to advance, several future actions are envisioned in relation to both antennas:

- *S-Band Patch Antenna*: The team has currently procured the Engineering and Flight Models of the S-Band Patch Antenna. Upcoming plans include further testing in an anechoic chamber to replicate results from the in-house development model tests. Apart from the gain and polarization characterization, we will conduct a link test between the spacecraft patch and the respective ground station feed to minimize polarization losses.

- *UHF Turnstile Deployable Antenna*: Numerous actions are planned for both the mechanism and the RF components of the UHF Turnstile Deployable Antenna. As elaborated in more detail in subsection 3.4, identified inefficiencies from environmental tests will be addressed in the next iteration, along with other ongoing improvement discussions. In order to increase our confidence to the mechanism, thermal simulations and structural analysis (inertia matrix update) shall be rerun after the aforementioned improvements are incorporated. We intend to perform follow-up testing in a thermal chamber to enhance reliability during extreme cold conditions and resolve thermal insulation issues. Furthermore, while the mechanism has undergone comprehensive testing, we will conduct rigorous antenna characterization to complete the assessment.

These endeavors underline our commitment to continually enhance the AcubeSAT antenna system, ensuring it meets the mission requirements, while developing an open-source and cost effective communications system solution.

**Acknowledgements**

We thank all the colleagues of SpaceDot Team for their invaluable help and contribution during the development and testing of the antennas. Specifically we wish to thank Mr. Apostolos Spanakis-Misirlis and Mr. Parmenion Mavrikakis for their preliminary work on the UHF Deployable Antenna, as well as Mr. George Karathanasopoulos for his work on the early stages of the mechanism's mechanical design. We thank the ESA Education Office and "Fly Your Satellite! 3" Program for the expertise and feedback we received throughout the development process. We also thank them for giving us access to the CubeSat Support Facility at the ESEC-Galaxia, Belgium, and sponsoring our travel and accommodation in order to perform the environmental tests. Our warm gratitude extends to our Aristotle University of Thessaloniki for providing us with the proper infrastructure (lab & cleanroom), especially the Department of Electrical and Computer Engineering, and specifically Prof. Traianos Yioultsis for providing us access to the anechoic chamber utilized for characterizing the S-Band Patch Antenna. We also thank Prisma Electronics SA for catering to all our assembly and harness needs. We thank Xometry Europe for sponsoring the ADM parts. We wish to thank Mr. Konstantinos Kanavouras for his mentoring throughout the writing process of this paper and for providing technical and editorial review as did Mr. Dimitrios Nentidis, Mr. Nikolaos Strimpas, Mr. Ioannis Dimoulios, Mr. Georgios Sklavenitis and Mr. Konstantinos Kyriakos. Lastly, we thank our families and friends that supported us all the way by showing patience and understanding.

**Acronyms**

| | |
|---|---|
| **ADCS** | Attitude, Determination and Control Subsystem |
| **ADM** | Antenna Deployment Mechanism |
| **AIV** | Assembly, Integration and Verification |
| **AUTh** | Aristotle University of Thessaloniki |
| **CDR** | Critical Design Review |
| **CDS** | CubeSat Design Specification |
| **COMMS** | Communications |
| **CSS** | Coarse Sun Sensor |
| **DUT** | Device Under Test |
| **EPS** | Electrical Power System |
| **EQM** | Engineering Qualification Model |
| **ESA** | European Space Agency |
| **FM** | Flight Model |
| **GS** | Ground Station |
| **HPBW** | Half Power Beamwidth |
| **IC** | Integrated Circuit |
| **LoC** | Lab-on-a-Chip |
| **LTAN** | Local Time of the Ascending Node |
| **MMCX** | Micro-Miniature Coaxial |
| **OBC** | On Board Computer |
| **PCB** | Printed Circuit Board |
| **PEEK** | Polyether Ether Ketone |
| **RF** | Radio Freqency |
| **SMA** | Sub-Miniature version A |
| **TC** | Telecommand |
| **TM** | Telemetry |
| **TM&TC** | Telemetry & Telecommand |
| **TRP** | Temperature Reference Point |
| **TVAC** | Thermal Vacuum Chamber |
| **UHF** | Ultra High Frequency |
| **VHF** | Very High Frequency |